\documentclass[12pt,a4paper]{article}

\usepackage[left=5em]{geometry}
\usepackage{amsmath}
\usepackage{graphicx}

\title{New Interaction between Dark Energy and Dark Matter Changes Sign during Cosmological Evolution}
\author{Cheng-Yi Sun\footnote{cysun@mailis.gucas.ac.cn; ddscy@163.com}\
$^{,1}$\ and Rui-Hong Yue\footnote{yueruihong@nbu.edu.cn}\ $^{,2}$
\\
 {$^1$\small Institute of Modern Physics, Northwest University, Xian 710069, P.R.
 China.}\\
{$^2$\small Faculty of Science, Ningbo University, Ningbo 315211,
P.R. China.}}

\begin{document}
\maketitle
\begin{abstract}
It is found by Cai and Su that the interaction between dark energy
and cold dark matter is likely to change the sign during the
cosmological evolution. Motivated by this, we suggest a new form of
interaction  between dark energy and dark matter, which changes from
negative to positive as the expansion of our universe changes from
decelerated to accelerated. We find that the interacting model is
consistent with the second law of thermodynamics and the
observational constraints. And, we also discuss the unified
adiabatic-squared sound speed of the model.
\end{abstract}

\ \ \ \ PACS: 95.36.+x, 98.80.-k, 95.35+d

\ \ \ \ {\bf {Key words: }}{dark energy, dark matter, interaction}

\section{Introduction}
Increasing astronomical observations \cite{Supernova,WMAP,LSS} tell
us that the present universe is dominated by the so-called dark
energy (DE), which accounts for $\simeq70\%$ of the critical mass
density and has been pushing the universe into accelerated expansion
\cite{dark energy1,dark energy2}. The other main component in the
universe is cold dark matter (CDM), which accounts for $\simeq30\%$
of the critical mass density and behaves as the pressureless dust.
However, we have known little about the nature of dark energy and
dark matter so far. The simplest candidate for dark energy is
Einstein's cosmological constant, which can fit the observations
well so far. But, the cosmological constant is plagued with
well-known fine-tuning and cosmic coincidence problems.

To alleviate these problems, dynamical DE models have been
considered in the literature. The simplest one is scalar-field dark
energy models. So far, a wide variety of scalar-field dark energy
models have been proposed, such as quintessence \cite{quitessence},
phantom \cite{phantom}, $k$-essence \cite{kessence}, tachyon
\cite{tachyon}, quintom \cite{quintom}, hessence \cite{hessence},
etc. Other dynamical dark energy models include Chaplygin gas models
\cite{Chaplygin}, braneworld models \cite{braneworld}, holographic
models \cite{HDE}, agegraphic models \cite{ADE}, etc. A lot of
efforts have been made to explore the nature of dark energy.
Furthermore, since no known symmetry in nature prevents or
suppresses a nonminimal coupling between DE and CDM, there may exist
interactions between the two components. At the same time, from the
observation side, no piece of evidence has been so far presented
against such interactions. Indeed, possible interactions between the
two dark components have been discussed intensively in recent years.
It is found that a suitable interaction can help to alleviate the
coincidence problem \cite{IDE1}. Different interacting models of
dark energy have been investigated \cite{IDE2,IDE3}.

In the literature, the model with interaction between DE and CDM is
usually described by the Friedmann equation
\begin{equation}
  \label{FE}
  H^2=\frac{\kappa^2}{3}(\rho_m+\rho_d),\quad \kappa^2\equiv8\pi G.
\end{equation}
and the two conservation laws
\begin{align}
  \label{CLM}
  \dot{\rho}_m+3H\rho_m&=Q,\\
  \label{CLD}
  \dot{\rho}_d+3H(\rho_d+p_d)&=-Q,
\end{align}
where $Q$ denotes the phenomenological interaction term; $\rho_m$
and $\rho_d$ are the energy densities of CDM and DE respectively;
$p_d$ is the pressure density of DE; $H\equiv\dot{a}/a$ is the
Hubble parameter; $a$ is the scale factor in the
Friedmann-Robertson-Walker (FRW) metric; a dot denotes the
derivative with respect to the cosmic time $t$. Usually, three forms
of $Q$ are used
\begin{align}
  \label{Q1}Q_1&=3bH\rho_d,\\
  \label{Q2}Q_2&=3bH(\rho_d+\rho_m),\\
  \label{Q3}Q_3&=3bH\rho_m,
\end{align}
where $b$ is the coupling constant. Then positive $b$ means that DE
decays into CDM, while negative $b$ means CDM decays into DE. In the
cases of $Q=Q_1$ and $Q=Q_2$, negative $b$ would lead $\rho_m$ to be
negative in the far future. For negative $b$ in the case of $Q=Q_3$,
no such difficulty exists. But in Ref.\cite{0712.0565}, from the
thermodynamical view, it is argued that the second law of
thermodynamics strongly favors that DE decays into CDM, i.e. $b$ is
positive (see Ref.\cite{0811.0099} for a different view). So
generally $b$ is taken to be positive.

However, recently it was found that the observations may favor the
decaying of CDM into DE \cite{0811.0099,0702015,0905.0672}.
Particularly, in Ref.\cite{0912.1943}, in a way independent of
specific interacting forms the authors fitted the interaction term
$Q$ with observations. They found that $Q$ was likely to cross the
noninteracting line $(Q=0)$, namely the sign of interaction $Q$
changed, around $z=0.5$. This raises a remarkable challenge  to the
interacting models, since the usual phenomenological forms of
interaction, as shown in the last paragraph, do not change their
signs during the cosmological evolution. As noted in
\cite{0912.1943}, more general forms of interaction should be
considered.

In our paper, we are interested in proposing such a new form of
interaction. It is known that our universe changes from deceleration
to acceleration around $z=0.5$ \cite{0803.0982}. Thus the
interacting term proportional to ($\rho_d-\rho_m$) would change its
sign naturally around $z=0.5$. So we may assume the new form of
interaction to be
\begin{equation}
  \label{Q4}
  Q_4=3\sigma H(\rho_d-\alpha\rho_m),
\end{equation}
where $\sigma$ is the coupling constant and $\alpha$ is a positive
constant of order unity. For simplicity, we take $\alpha=1$. So the
new interaction term is assumed to be
\begin{equation}
  \label{nQ}
  Q_4=3\sigma H(\rho_d-\rho_m).
\end{equation}
Henceforth, we will denote the new interacting model as $\sigma$.
The parameter $\sigma$ is assumed to be positive, since negative
$\sigma$ would lead to negative $\rho_m$ in the far future.
Obviously, in the early stage, $Q_4$ is negative, since
$\rho_m>\rho_d$. As the expansion of our universe changes from
decelerated to accelerated, $Q_4$ changes from negative to positive.

Below, we first show that the interacting model with $Q=Q_4$ is
consistent with the second law of thermodynamics by following the
argument in \cite{0712.0565}. Second, we calculate the unified
adiabatic sound speeds of the interacting model.  Then we compare
the model with observations. Finally, conclusions are given.

\section{New Interaction and The Second Law of Thermodynamics}

We recall the thermodynamical description of DE and CDM in
\cite{0712.0565,0811.0099,9609119}. A perfect fluid is characterized
by ($n,\rho,p,s,u^a$), where $n$ is the particle number density,
$\rho$ is the energy density, $p$ is the pressure density, $s$ is
the entropy per particle and $u^a$ is the 4-velocity. In the FRW
universe, we take $u^a=(\partial/\partial t)^a$ and ${u^a}_{;a}=3H$.
In this paper, a semicolon denotes the covariant derivative
compatible with the FRW metric. The energy-momentum tensor may be
assumed to be \cite{9609119}
\begin{equation}
  \label{EMTensor}
  T^{ab}=\rho u^au^b+(p+\Pi)(g^{ab}+u^au^b).
\end{equation}
Then the conservation law of the fluid $u_a{T^{ab}}_{;b}$=0 gives us
\cite{0811.0099}
\begin{equation}
  \label{CLPerfectF}
  \dot{\rho}+3H(\rho+p)=-3H\Pi,
\end{equation}
where $-3H\Pi$ represents the phenomenological interaction between
the fluid and others. The balance equation for the particle number
is assumed to be \cite{0811.0099}
\begin{equation}
  \label{CLNumber}
  \dot{n}+3Hn=n\Gamma,
\end{equation}
where $\Gamma$ is the rate of the change of the particle number of
the fluid. The temperature $T$ of the fluid is defined via the Gibbs
equation
\begin{equation}
  \label{GibbsEq}
  Tds=d\Big(\frac{\rho}{n}\Big)+pd\Big(\frac{1}{n}\Big),
\end{equation}
so that the variation rate of the entropy per particle is \cite{0811.0099}
\begin{equation}
  \label{EntropyRate}
  \dot{s}=-\frac{3H\Pi}{Tn}-\frac{\rho+p}{Tn}\Gamma.
\end{equation}
By defining the entropy flow vector as
\begin{equation}
  \label{EntropyVector}
  S^a=snu^a,
\end{equation}
we have \cite{0811.0099}
\begin{equation}
  \label{DivEntropy}
  {S^a}_{;a}=(s-\frac{\rho+p}{nT})n\Gamma-\frac{3H\Pi}{T}.
\end{equation}
The entropy per particle is \cite{0811.0099}
\begin{equation}
  \label{EntropyParicle}
  s=\frac{\rho+p}{nT}-\frac{\mu}{T},
\end{equation}
where $\mu$ is the chemical potential. Then, we have
\begin{equation}
  \label{DivEntropy1}
  {S^a}_{;a}=-\frac{\mu}{T}n\Gamma-\frac{3H\Pi}{T}.
\end{equation}

In the general case, the Gibbs equation can be rewritten as
\begin{equation}
  \label{GibbsEq2}
  ds=-\frac{\rho+p}{Tn^2}dn+\frac{1}{Tn}d\rho,
\end{equation}
and the integrability condition
\[
  \frac{\partial^2s}{\partial \rho\partial n}=\frac{\partial^2s}{\partial n\partial\rho}
\]
tells us
\begin{equation}
  \label{intCondtion}
  n\frac{\partial T}{\partial n}+(\rho+p)\frac{\partial T}{\partial\rho}=T\frac{\partial
  p}{\partial\rho}.
\end{equation}
Using Eqs.(\ref{CLPerfectF}), (\ref{CLNumber}), (\ref{EntropyRate})
and (\ref{intCondtion}), we have
\begin{equation}
  \label{dTdtT}
  \frac{\dot{T}}{T}=-3H\frac{\partial p}{\partial\rho}+n\dot{s}\frac{\partial T}{\partial\rho}+\Gamma\frac{\partial p}{\partial\rho}
\end{equation}
Generally, we should assume that at any event in spacetime, the
thermodynamics-state of the fluid is close to the fictitious
equilibrium-state at that event \cite{9609119}. This implies that
the right-hand side of Eq.(\ref{dTdtT}) is dominated by the first
term, and we have approximately
\begin{equation}
  \label{dTdtTAppr}
  \frac{\dot{T}}{T}\simeq-3H\frac{\partial p}{\partial\rho}
\end{equation}

Now we apply the results above to the model with interaction between
DE and CDM. Usually the energy-momentum tensor (EMT) of CDM is taken
as
\begin{equation}
  \label{EMTCDM}
  T_\text{m}^{ab}=\rho_m u_1^au_1^b+p_m(g^{ab}+u_1^au_1^b).
\end{equation}
Then the conservation law should be
\begin{equation}
  \label{CLEMTCDM}
  {T_{\text{m}}^{ab}}_{;b}=[\Pi(g^{ab}+u_1^au_1^b)]_{;b}.
\end{equation}
Here $u_1^a$ is the four-velocity of CDM. In this paper, we use the
subscripts $m$ and $d$ to denote the corresponding parameters of CDM
and DE, respectively. With $p_m=0$ and choosing
\begin{equation}
  \label{Pi}
  \Pi=\sigma(\rho_d-\rho_m),
\end{equation}
we can deduce Eq.(\ref{CLM}) with $Q=Q_4$ by contracting
Eq.(\ref{CLEMTCDM}) with $u_{1a}$. Then equivalently we can define
the effective EMT of CDM as
\begin{equation}
  \label{EMTCDMEff}
  T_{\text{me}}^{ab}=\rho_m u_1^au_1^b+(p_m-\Pi)(g^{ab}+u_1^au_1^b).
\end{equation}
Obviously, the effective EMT of CDM is conserved
\[
  {T_{\text{me}}^{ab}}_{;b}=0.
\]
Similarly, although the EMT of DE
\[
  T_\text{d}^{ab}=\rho_d u_2^au_2^b+p_d(g^{ab}+u_2^au_2^b),
\]
is not conserved
\[
  {T_{\text{d}}^{ab}}_{;b}=-[\Pi(g^{ab}+u_2^au_2^b)]_{;b},
\]
we can define the effective EMT of DE as
\begin{equation}
  \label{EMTDEEff}
  T_{\text{de}}^{ab}=\rho_d u_2^au_2^b+(p_d+\Pi)(g^{ab}+u_2^au_2^b),
\end{equation}
which is also conserved
\[
  {T_{\text{de}}^{ab}}_{;b}=0.
\]
Here $u_2^a$ is the four-velocity of DE. We can recover
Eq.(\ref{CLD}) with $Q=Q_4$ from $u_{2a}{T_{\text{de}}^{ab}}_{;b}=0$
if $\Pi$ is chosen as given in Eq.(\ref{Pi}).

The equation of state of DE is
\begin{equation}
  \label{w}
  p_d=w\rho_d.
\end{equation}
In this paper, we only consider the model with constant $w$. For
CDM, approximately we have \cite{9609119}
\begin{equation}
  \label{EoSCDM}
  \rho_m=n_mM+\frac{3}{2}n_mT_m,\quad p_m=n_mT_m \quad (k_B=1),
\end{equation}
so long as $T_m\ll M$. From Eq.(\ref{dTdtTAppr}), approximately we
have
\begin{equation}
  \label{TA}
  T_m\propto a^{-2},\quad T_d\propto a^{-3w}.
\end{equation}
The results tell us that as the universe expands, the temperature of
CDM, $T_m$, decreases and the temperature of DE, $T_d$, increases.
Thus, one may expect that at the present and in the future
$T_m<T_d$, while in the past $T_m>T_d$. Following
Ref.\cite{0712.0565}, we assume that both DE and CDM have
null-chemical potentials. Thus from Eq.(\ref{DivEntropy1}), we have
\begin{equation}
  \label{DivEntropyTot}
  S^a_{m;a}+S^a_{d;a}=\Big(\frac{1}{T_m}-\frac{1}{T_d}\Big)Q,
\end{equation}

The existence of interaction means that there is a transfer of
energy between DE and CDM. It is a natural conclusion that nowadays
the energy is transferred from DE to CDM, i.e. $Q>0$, since
currently $T_d>T_m$. In addition, from Eq.(\ref{DivEntropyTot}), the
second law of thermodynamics $S^a_{m;a}+S^a_{d;a}\ge0$ and because
$T_m<T_d$ indicate that currently $Q>0$.

The interaction term $Q=Q_1$ is used in the analysis in
\cite{0712.0565}; $Q_1$ cannot change its sign, and is still
positive even when $T_m>T_d$. It seems that at earlier, the second
law of thermodynamics was violated, since $T_m>T_d$, $Q_1>0$ and
Eq.(\ref{DivEntropyTot}) indicate that $S^a_{m;a}+S^a_{d;a}<0$ and
the energy is being transferred from the lower temperature DE to the
higher temperature CDM. To overcome the difficulty, the author in
\cite{0712.0565} argued that the thermodynamical description
breaks-down at some point, both when $a\ll1$ and when $a\gg1$.

Now, we apply the analysis in \cite{0712.0565} to the case of
$Q=Q_4$ in Eq.(\ref{nQ}). Obviously, nowadays the second law of
thermodynamics is satisfied since $Q_4$ is positive. Earlier, since
$\rho_m>\rho_d$, $Q_4$ was negative, which indicates the energy is
transferred from CDM to DE. In fact, this is just  what is expected
from the second law of thermodynamics when $T_m>T_d$. So, the
interacting model $\sigma$ with $Q=Q_4$ is always consistent with
the second law of thermodynamics even at early times.

\section{Adiabatic Sound Speed}
\label{SecASS}

The squared sound speed $c_s^2$, defined as
\[
c_s^2=\frac{\delta p}{\delta\rho},
\]
is an important quantity for the cosmological evolution, which
determines the stability of the cosmological evolution
\cite{0207347}. The adiabatic-squared sound speed $c_a^2$ is defined
as
\begin{equation}
  \label{ASS}
  c_a^2=\frac{\dot{p}}{\dot{\rho}}.
\end{equation}
In the interacting model $\sigma$, we rewrite Eqs.(\ref{CLM}) and
(\ref{CLD}), respectively as
\begin{align}
  \label{CLMeff}
  \dot{\rho}_m+3H(\rho_m+p^{\text{eff}}_m)&=0,\\
  \label{CLDeff}
  \dot{\rho}_d+3H(\rho_d+p^{\text{eff}}_d)&=0,
\end{align}
where
\begin{align}
  \label{pmEff}
  p^{\text{eff}}_m&=-\sigma(\rho_d-\rho_m),\\
  \label{pdEff}
  p^{\text{eff}}_d&=p_d+\sigma(\rho_d-\rho_m).
\end{align}
Naively, we may still define the squared sound speed of DE
$c_{sd}^2$ as
\begin{equation}
  \label{csd}
  c^2_{sd}=\frac{\delta p_d}{\delta\rho_d}.
\end{equation}
Actually, this is not the physical sound speed of DE. (The two
physical sound speeds $\lambda_\pm$ are shown in the Appendix.)
However, we find that in the model $\sigma$ the stabilities of CDM
and DE under perturbations are still determined by the squared sound
speed of DE $c_{sd}^2$ (see the Appendix for details). The
corresponding adiabatic sound speed is
\begin{equation}
  \label{cat}
  c^2_{ad}=\frac{\dot{p}_d}{\dot{\rho}_d}.
\end{equation}
Then, using Eqs.(\ref{w}), we have
\begin{equation}
  \label{catW}
  c^2_{ad}=w.
\end{equation}
Negative $w$ indicates that adiabatic instabilities exist. Yet, this
is not very astonishing since it is well-known that even in the
noninteracting model with constant $w$, the adiabatic squared sound
speed of DE $c_{ad}^2$ is negative ($c_{ad}^2=w<0$), leading to the
adiabatic instabilities.

\section{Comparison with Observational Data}

In this section, first we will explore whether the interacting model
$\sigma$ is consistent with the results in Ref.\cite{0710.5345}.
Second, we will explore whether the model $\sigma$ is consistent
with the observational constraint on the position of the first peak
of the cosmic microwave background power spectrum.

In order to explore whether the model $\sigma$ is consistent with
the results in \cite{0710.5345}, we should calculate the
dimensionless coordinate distance, $y(z)=H_0a_0\tilde{r}$, and the
two first-derivatives with respect to redshift, and then compare the
results with the observational data of supernovae-type Ia (SN Ia)
and radio-galaxies between the redshift $z=0$ and $z=1.8$. Here
$z\equiv\frac{a_0}{a}-1$ is the cosmological redshift and
$\tilde{r}$ is the radial coordinate in the FRW metric.

Since $dt=-a(t)d\tilde{r}$ for photons flying from their sources to
observer in the flat-FRW universe, we have
\begin{equation}
  \label{EY}
  E(z)\equiv\frac{H(z)}{H_0}=\frac{1}{y'(z)},
\end{equation}
where $y'\equiv dy/dz$. We define
\begin{equation}
  \label{rhoIm}
  \rho_m=\rho_{m0}\Im_1(z),\quad \rho_d=\rho_{d0}\Im_2(z).
\end{equation}
Then using Eqs.(\ref{FE}), (\ref{CLM}), (\ref{CLD}), (\ref{EY}) and
(\ref{rhoIm}), we have
\begin{align}
  \label{yp}
  y'(z)&=\frac{1}{[\Omega_{m0}\Im_1+\Omega_{d0}\Im_2]^{1/2}},\\
  \label{ypp}
  y''(z)&=-\frac{3}{2}\frac{y'}{1+z}\Big[\frac{\Omega_{m0}\Im_1+(1+w)\Omega_{d0}\Im_2}{\Omega_{m0}\Im_1+\Omega_{d0}\Im_2}\Big],
\end{align}
where $y''\equiv d^2y/dz^2$.

Now let us calculate $\Im_1$ and $\Im_2$. By using Eqs.(\ref{nQ}),
(\ref{rhoIm}) and (\ref{w}), the conservation laws (\ref{CLM}) and
(\ref{CLD}) read respectively
\begin{align}
  \label{dIm1da}
  \frac{d\Im_1}{da}+\frac{3(1+\sigma)}{a}\Im_1&=\frac{3\sigma}{ar_0}\Im_2,\\
  \label{dIm2da}
  \frac{d\Im_2}{da}+\frac{3(1+w+\sigma)}{a}\Im_2&=\frac{3r_0\sigma}{a}\Im_1,
\end{align}
where $r_0\equiv\frac{\rho_{m0}}{\rho_{d0}}$. By solving the two
equations, we have
\begin{align}
  \label{Im1}
  \Im_1(z)&=c(1+z)^{s_+}+(1-c)(1+z)^{s_-},\\
  \label{Im2}
  \Im_2(z)&=-\frac{r_0}{3\sigma}[c\tilde{s}_+(1+z)^{s_+}+(1-c)\tilde{s}_-(1+z)^{s_-}],
\end{align}
where
\begin{align}
  \label{spm}
  s_\pm&=\frac{3}{2}[2(1+\sigma)+w\pm\sqrt{4\sigma^2+w^2}],\\
  \label{tildespm}
  \tilde{s}_\pm&=\frac{3}{2}(w\pm\sqrt{4\sigma^2+w^2}),\\
  \label{c}
  c&=\frac{1}{2}-\frac{w+\frac{2\sigma}{r_0}}{2\sqrt{w^2+4\sigma^2}}.
\end{align}
In the absence of interaction, the Eqs. (\ref{yp}) and (\ref{ypp})
reduce to Eqs.(1) and (2) of Ref.\cite{0710.5345}.

\begin{figure}
\centering
\renewcommand{\figurename}{Fig.}
\includegraphics[scale=0.6]{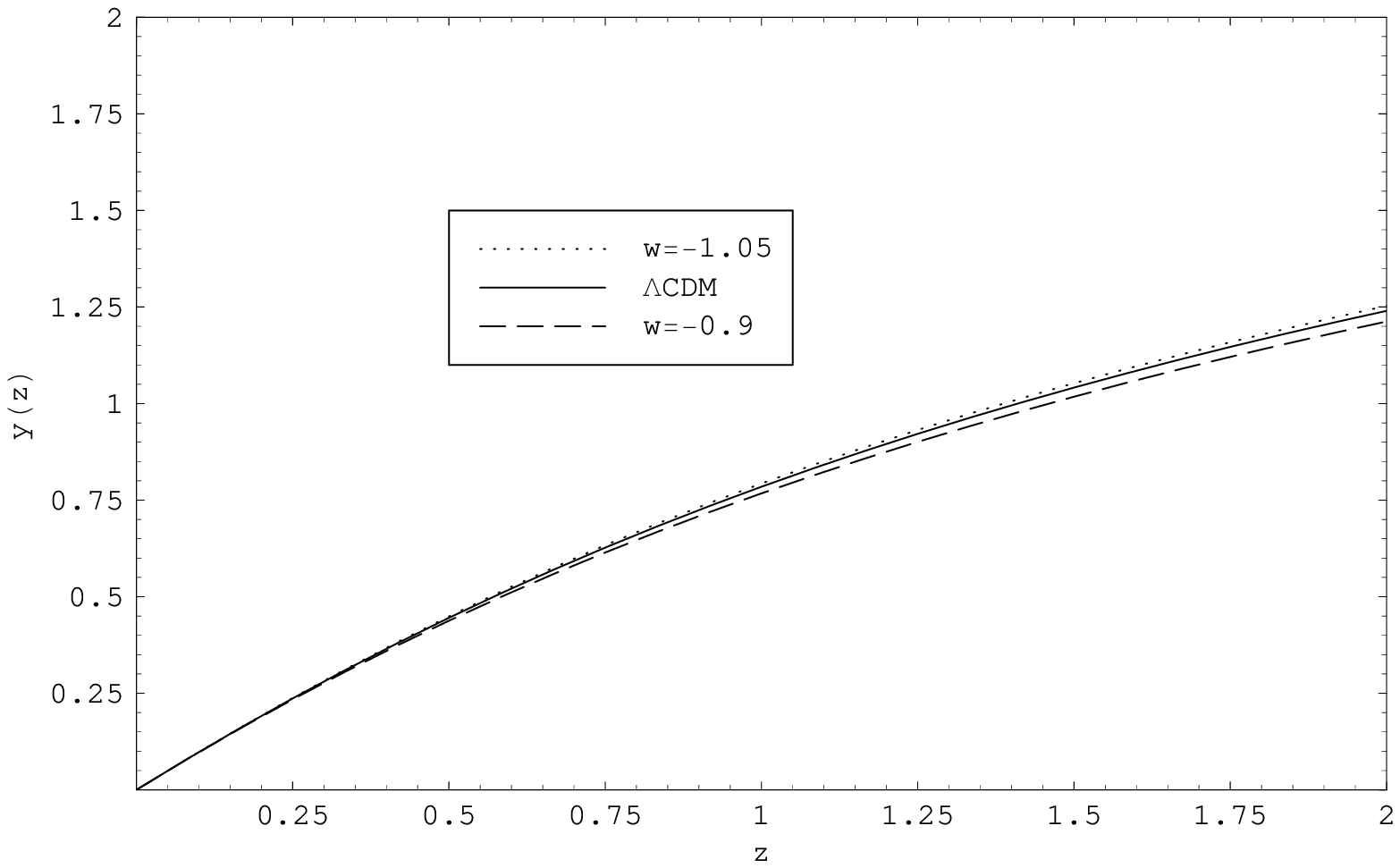}
\caption{$y(z)$ versus $z$ in the interacting model $\sigma$ for
fixed $\Omega_{m0}=0.272$, $\sigma=10^{-3}$ and different $w$. For
comparison, the prediction of the $\Lambda$CDM model is also shown.
\label{FigY}}
\end{figure}

\begin{figure}
\centering
\renewcommand{\figurename}{Fig.}
\includegraphics[scale=0.6]{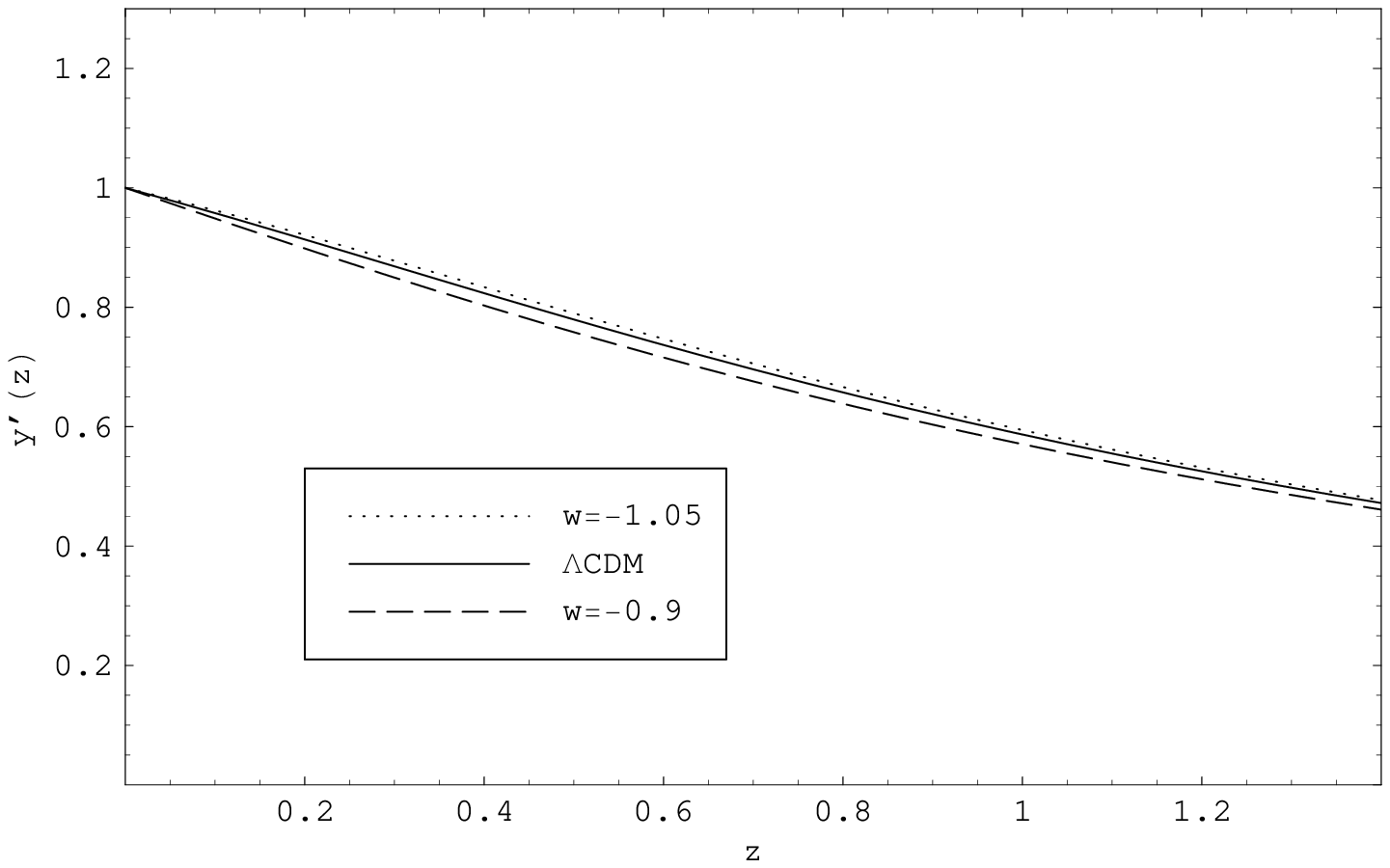}
\caption{$y'(z)$ versus $z$ in the interacting model $\sigma$ for
fixed $\Omega_{m0}=0.272$, $\sigma=10^{-3}$ and different $w$. For
comparison, the prediction of the $\Lambda$CDM model is also shown.
\label{FigYp}}
\end{figure}

\begin{figure}
\centering
\renewcommand{\figurename}{Fig.}
\includegraphics[scale=0.6]{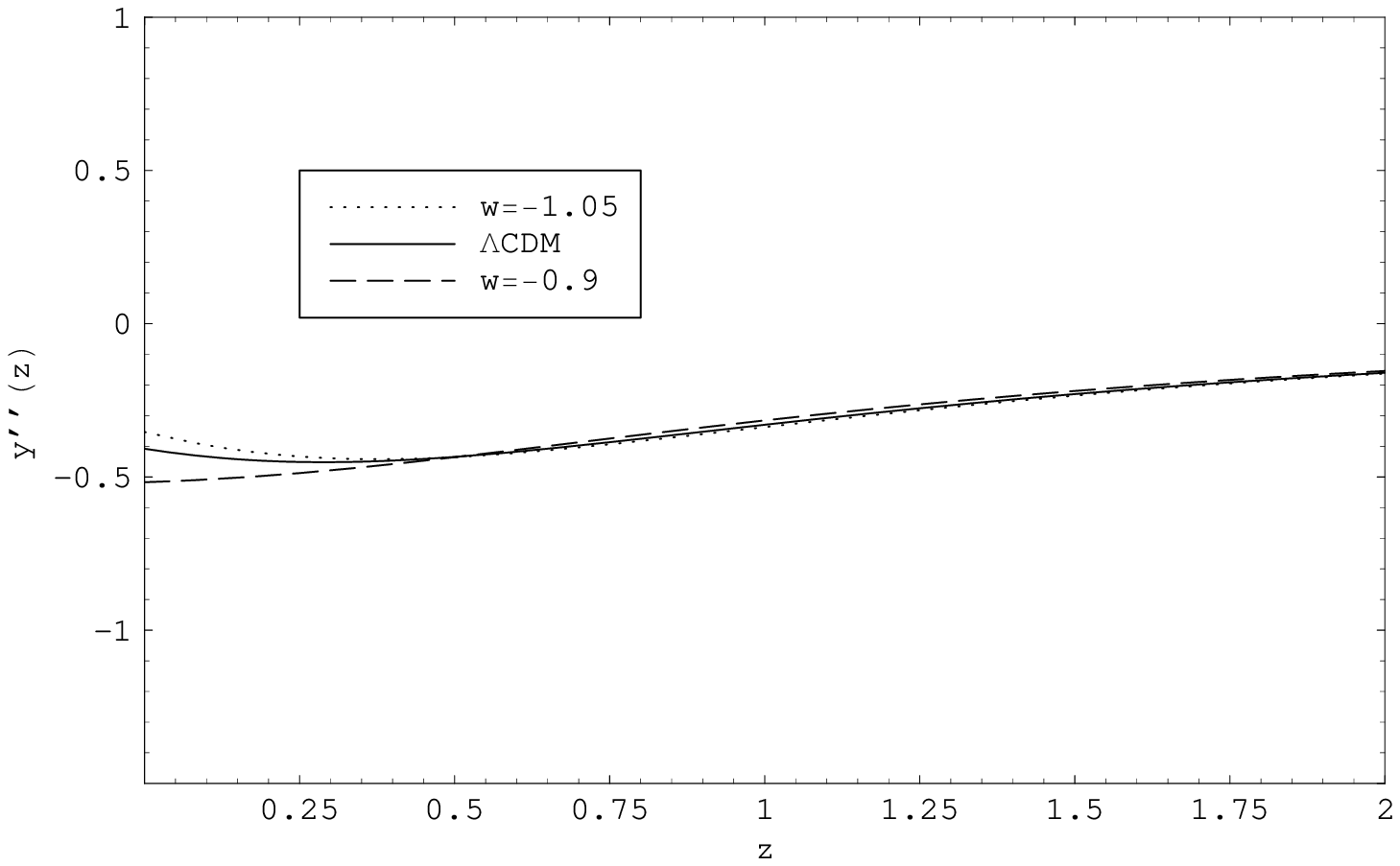}
\caption{$y''(z)$ versus $z$ in the interacting model $\sigma$ for
fixed $\Omega_{m0}=0.272$, $\sigma=10^{-3}$ and different $w$. For
comparison, the prediction of the $\Lambda$CDM Model is shown, too.
\label{FigYpp}}
\end{figure}

The evolutions of $y$, $y'$ and $y''$ with respect to $z$ are
depicted in Fig.\ref{FigY}, Fig.\ref{FigYp} and Fig.\ref{FigYpp}. We
have used the different values of the equation of state parameter
$w$, and fixed the values of $\Omega_{m0}=0.272$ \cite{WMAP7Year}
and $\sigma=10^{-3}$. Comparisons of Figs.\ref{FigY},
Fig.\ref{FigYp} and Fig.\ref{FigYpp} with the corresponding figures
in Ref.\cite{0710.5345} reveal that the interacting model $\sigma$
is consistent with the analysis of Ref.\cite{0710.5345}.

Now, let us calculate the shift parameter $R$, which characterizes
the position of the first peak of the cosmic microwave background
spectrum and is defined as \cite{shiftPara}
\begin{equation}
  \label{shiftParas}
  R=\sqrt{\Omega_m}\int^{z_*}_0{\frac{dz}{E(z)}}.
\end{equation}
Here $z_*$ is the redshift of decoupling. The 7-year Wilkinson
Microwave Anisotropy Probe(WMAP) observations tell us that
$z_*=1091.3\pm0.91$ at $1\sigma$ confidence level \cite{WMAP7Year}.
In this paper, we fix $z_*=1091$. The values of the shift parameter
$R$ for different sets of parameters in the model $\sigma$ and the
$\Lambda$CDM model are displayed in Table \ref{Tab1}. We have fixed
$\Omega_{m0}=0.272$ and used the flat FRW metric. We first consider
that the universe is filled with DE ($\Omega_{d0}=0.728$) and CDM
($\Omega_{m0}=0.272$). Second, we consider the case that the
universe is filled with DE ($\Omega_{d0}=0.728$), CDM
($\Omega_{dm0}=0.2264$), and baryon matter ($\Omega_{b0}=0.0456$)
\cite{WMAP7Year}. The total fractional-energy density of CDM and
baryon matter is still fixed to be $\Omega_{m0}=0.272$. The
component of baryon matter is assumed to be evolving separately. In
the $\Lambda$CDM model, the shift parameter $R$ is determined by the
total fractional energy density of CDM and baryons. The 7-year WMAP
observations tell us $R=1.725\pm0.018$ at $1\sigma$ confidence
level. Then from the results displayed in Table \ref{Tab1}, we know
that the interacting model $\sigma$ is consistent with the 7-year
WMAP observations.

\begin{table}
\centering
\renewcommand{\tablename}{TABLE}
\begin{tabular}{c|cccc|c} \hline Model & $\Omega_{m0}$ & $\Omega_{b0}$ & $w$ & $\sigma$ & R \\
\hline  $\sigma$      & $0.272$ & $0$ &$-0.9$ & $10^{-3}$ & $1.712$\\
\hline  $\sigma$      & $0.272$ & $0$ &$-1.05$ & $10^{-3}$ & $1.738$\\
\hline  $\sigma$      & $0.272$ & $0.0456$ &$-0.9$ & $10^{-3}$ & $1.713$\\
\hline  $\sigma$      & $0.272$ & $0.0456$ &$-1.05$ & $10^{-3}$ & $1.739$\\
\hline $\Lambda$CDM   & $0.272$ &   &$-1$  & $10^{-3}$  & $1.733$\\
\hline
\end{tabular}
\caption{The values of the shift parameter for different sets of
parameters in the model $\sigma$ and  the $\Lambda$CDM
model.\label{Tab1}}
\end{table}

\section{Conclusions}

The dark energy models with interaction between DE and CDM have been
investigated intensively. It is argued in \cite{0812.2210} that the
interacting models of dark energy may be key to solving the cosmic
coincidence problem. Recently, it was found that the interaction is
likely to cross the noninteracting line \cite{0912.1943}. However,
the usual forms of interaction used in the literature can not change
their sign. The result in \cite{0912.1943} raises a challenge to the
interacting models of dark energy; more general interaction is
needed to be considered. In \cite{1008.4968}, the interacting models
with interaction proportional to the deceleration parameter are
suggested.

In this paper, we suggest the new interacting model $\sigma$ with
interaction $Q=Q_4$. Obviously the sign of $Q_4$ changes from
negative to positive as the expansion of our universe changes from
decelerated to accelerated. We found that the interacting model is
consistent with the second law of thermodynamics. Then we also found
the squared sound speed of DE, $c_{sd}^2$, to be crucial in
determining the stabilities of DE and CDM in the model. As in the
noninteracting model with constant $w$, there also exist adiabatic
instabilities in the model $\sigma$ with constant $w$ due to
$c_{ad}^2=w<0$. Furthermore, we compared the interacting model with
observational data. And we found the interaction model to be
consistent with the result in \cite{0710.5345} and the 7-year WMAP
observations \cite{WMAP7Year}. Thus, we believe that the interacting
model with $Q=Q_4$ is consistent with observational constraints. In
the future, we plan to explore how to obtain the phenomenological
interaction term $Q_4$ from an action principle.

\section*{Acknowledgments}
This work has been supported in part by the National Natural Science
Foundation of China under Grants No. 10875060 and No. 11147017, the
Natural Science Foundation of the Northwest University of China
under Grant No. 09NW27, and the Research Fund for the Doctoral
Program of Higher Education of China under Grant No. 20106101120023.

\section*{Appendix}
Here, we try to show that the squared sound speed of DE determines
the stabilities of the cosmological evolution in the interacting
model. We adopt the conformal Newtonian gauge. Then the perturbed
metric about a sptially flat, homogeneous and isotropic FRW
background is given to be \cite{Mukhanov}
\begin{equation}
  \label{metric}
  ds^2=a^2[-(1+2\phi)d\tau^2+(1-2\psi)d\textbf{x}^2].,
\end{equation}
where a is the scale factor and the perturbations of the metric are
characterized by two potentials, $\phi$ and $\psi$. The total EMT is
\begin{equation}
  \label{EMTTotal}
  T_{\text{t}}^{ab}=T_{\text{me}}^{ab}+T_{\text{de}}^{ab}.
\end{equation}
The $T_{\text{me}}^{ab}$ and $T_{\text{de}}^{ab}$ are the effective
EMT of CDM and DE, respectively,
\begin{align}
  \label{EMTCDMEff2}
  T_{\text{me}}^{ab}&=\rho_mu_1^au_1^b+p_m^{\text{eff}}(u_1^au_1^b+g^{ab}),\\
  \label{EMTDEEff2}
  T_{\text{de}}^{ab}&=\rho_du_2^au_2^b+p_d^{\text{eff}}(u_2^au_2^b+g^{ab}).
\end{align}
Here $p_m^{\text{eff}}$ and $p_d^{\text{eff}}$ are defined in
Eqs.(\ref{pmEff}) and (\ref{pdEff}) respectively, and
${T_{\text{me}}^{ab}}_{;b}={T_{\text{me}}^{ab}}_{;b}=0$. To the
first-order of perturbation, we have
\begin{align}
  \label{PertRhom}
  \rho_m(\tau,\textbf{x})&=\rho_{mb}(\tau)[1+\delta_1(\tau,\textbf{x})],\\
  \label{PertPm}
  p^{\text{eff}}_m(\tau,\textbf{x})&=p^{\text{eff}}_{mb}(\tau)+\delta p^{\text{eff}}_m(\tau,\textbf{x}),\\
  \label{PertVelo}
  u_1^a(\tau,\textbf{x})&=a^{-1}[(1-\phi)(\partial/\partial \tau)^a+\partial_iv_1(\tau,\textbf{x})(\partial/\partial
  x^i)^a],
\end{align}
and
\begin{align}
  \label{PertRhod}
  \rho_d(\tau,\textbf{x})&=\rho_{db}(\tau)[1+\delta_2(\tau,\textbf{x})],\\
  \label{PertPm}
  p^{\text{eff}}_d(\tau,\textbf{x})&=p^{\text{eff}}_{db}(\tau)+\delta p^{\text{eff}}_d(\tau,\textbf{x}),\\
  \label{PertVelo}
  u_2^a(t,\textbf{x})&=a^{-1}[(1-\phi)(\partial/\partial\tau)^a+\partial_iv_2(\tau,\textbf{x})(\partial/\partial
  x^i)^a].
\end{align}
Here we use the subscript $b$ to denote the spatially
homogeneous-background value of the corresponding quantity, and
$\delta_1\equiv\frac{\delta\rho_m}{\rho_{mb}}$ and
$\delta_2\equiv\frac{\delta\rho_d}{\rho_{db}}$ are the fractional
perturbations in the energy densities of CDM and DE, respectively;
$v_1$ and $v_2$ are the peculiar velocity potentials of CDM and DE,
respectively, with the same order of $\delta_1$ and $\delta_2$;
$\delta p^{\text{eff}}_m(t,\textbf{x})$ and $\delta
p^{\text{eff}}_d(t,\textbf{x})$ are the perturbations of
$p^{\text{eff}}_m(t,\textbf{x})$ and
$p^{\text{eff}}_d(t,\textbf{x})$, respectively,
\begin{align}
  \label{deltaPm}
  \delta p^{\text{eff}}_m(\tau,\textbf{x})&=-\sigma[\rho_{db}(\tau)\delta_2(\tau,\textbf{x})-\rho_{mb}(\tau)\delta_1(\tau,\textbf{x})],\\
  \label{deltaPd}
  \delta p^{\text{eff}}_d(\tau,\textbf{x})&=\delta
  p_d(\tau,\textbf{x})+\sigma[\rho_{db}(\tau)\delta_2(\tau,\textbf{x})-\rho_{mb}(\tau)\delta_1(\tau,\textbf{x})],
\end{align}
where $\delta
p_d(\tau,\textbf{x})=p_d(\tau,\textbf{x})-p_{db}(\tau)$.

In the conformal Newtonian gauge, the first-order perturbed Einstein
equations give us \cite{9506072}
\begin{align}
  \label{EinsteinEq00}
  &3\mathcal{H}\psi'+k^2\psi+3\mathcal{H}^2\phi=-4\pi Ga^2(\delta_1\rho_{mb}+\delta_2\rho_{db}),\\
  \label{EinsteinEq0i}
  &k^2\psi'+k^2\mathcal{H}\phi=4\pi Ga^2[(\rho_{mb}+p^{\text{eff}}_{mb})\theta_1+(\rho_{db}+p^{\text{eff}}_{db})\theta_2]\\
  \label{EinsteinEqii}
  &\psi''+\mathcal{H}(2\psi'+\phi')+(2\frac{a''}{a}-\mathcal{H}^2)\phi+\frac{k^2}{3}(\psi-\phi)=4\pi Ga^2\delta p_d\\
  \label{EinsteinEqij}
  &\psi-\phi=0.
\end{align}
Hereafter, primes denote the derivatives with respective to the
conformal time $\tau$, $\mathcal{H}'\equiv a'/a$,
$\theta_1\equiv-k^2v_1$ and $\theta_2\equiv-k^2v_2$. The first-order
equation of the conservation law of CDM ${T_{me}^{ab}}_{;b}=0$ (in
Fourier space) tells us \cite{9506072}
\begin{align}
  \label{CLMeff0}
  \delta'_1-3\mathcal{H}\frac{p^{\text{eff}}_{mb}}{\rho_{mb}}\delta_1+3\mathcal{H}\frac{\delta p^{\text{eff}}_m}{\rho_{mb}}
                        -3(1+\frac{p^{\text{eff}}_{mb}}{\rho_{mb}})\psi'+(1+\frac{p^{\text{eff}}_{mb}}{\rho_{mb}})\theta_1&=0,\\
  \label{CLMeffi}
  \theta_1'+\frac{\rho'_{mb}+p'^{\text{eff}}_{mb}}{\rho_{mb}+p^{\text{eff}}_{mb}}\theta_1
                        +4\mathcal{H}\theta_1-k^2\phi-k^2\frac{\delta p^{\text{eff}}_m}{\rho_{mb}+p^{\text{eff}}_{mb}}&=0.
\end{align}
The first-order equation of the conservation law of DE
${T_{de}^{ab}}_{;b}=0$ tells us that \cite{9506072}
\begin{align}
  \label{CLDeff0}
  \delta'_2-3\mathcal{H}\frac{p^{\text{eff}}_{db}}{\rho_{db}}\delta_2+3\mathcal{H}\frac{\delta p^{\text{eff}}_d}{\rho_{bb}}
                        -3(1+\frac{p^{\text{eff}}_{db}}{\rho_{db}})\psi'+(1+\frac{p^{\text{eff}}_{db}}{\rho_{db}})\theta_2&=0,\\
  \label{CLDeffi}
  \theta_2'+\frac{\rho'_{db}+p'^{\text{eff}}_{db}}{\rho_{db}+p^{\text{eff}}_{db}}\theta_2
                        +4\mathcal{H}\theta_2-k^2\phi-k^2\frac{\delta p^{\text{eff}}_d}{\rho_{db}+p^{\text{eff}}_{db}}&=0.
\end{align}
By differentiating Eq.(\ref{CLMeff0}) with respect to $\tau$ and
using Eqs(\ref{EinsteinEqii}), (\ref{EinsteinEqij}), (\ref{CLMeffi})
and (\ref{EinsteinEq00}), taking the geometric optic limit, finally
we can get
\begin{equation}
  \label{ddDelta1DtauAppr}
  \delta''_1=-k^2\frac{\delta p^{\text{eff}}_{m}}{\rho_{mb}}.
\end{equation}
Similarly, from Eq.(\ref{CLDeff0}), and using
Eqs(\ref{EinsteinEqii}), (\ref{EinsteinEqij}), (\ref{CLDeffi}) and
(\ref{EinsteinEq00}), taking the geometric optic limit, we can also
get
\begin{equation}
  \label{ddDelta2DtauAppr}
  \delta''_2=-k^2\frac{\delta p^{\text{eff}}_{d}}{\rho_{db}}.
\end{equation}
By using Eqs.(\ref{deltaPm}) and (\ref{deltaPd}), we can rewrite the
above two equations as
\begin{equation}
  \label{ddDeltaDtauMatrix}
  \frac{d^2}{d\tau^2}\begin{pmatrix}\delta_1\\
  \delta_2\end{pmatrix}=-k^2\mathcal{M}\begin{pmatrix}\delta_1\\
  \delta_2\end{pmatrix}.
\end{equation}
Here the matrix $\mathcal{M}$ is
\begin{equation}
  \label{matrixM}
  \mathcal{M}=
  \begin{pmatrix}
    \sigma        & -\sigma r\\
    -\sigma r^{-1} & c_{sd}^2+\sigma
  \end{pmatrix},
\end{equation}
where
\[
  r\equiv\frac{\rho_{db}}{\rho_{mb}},\quad c_{sd}^2\equiv\frac{\delta p_{d}}{\delta\rho_{d}}.
\]
This matrix possesses two eigenvalues $\lambda_+$ and $\lambda_-$,
\begin{equation}
  \label{lambdapm}
  \lambda_\pm=\frac{c_{sd}^2+2\sigma\pm\sqrt{c_{sd}^4+4\sigma^2}}{2}.
\end{equation}
Clearly, $\lambda_\pm$ are crucial for determining the stability of
the interacting model. In fact, $\lambda_\pm$ are the physical sound
speeds in the interacting model. If $\lambda_+\geq0$ and
$\lambda_-\ge0$, the model is stable. Otherwise, the model is
instable. Here we assume $\sigma\ge0$ and it can be easily checked
that $c_{sd}^2>0$ indicates $\lambda_\pm>0$ and $c_{sd}^2<0$
indicates $\lambda_-<0$. So we know $c^2_{sd}$ is crucial for
determining the stability of the interacting model, and negative
$c^2_{sd}$ indicates the existence of instabilities.

\end{document}